# EffiTest: Efficient Delay Test and Statistical Prediction for Configuring Post-silicon Tunable Buffers

Grace Li Zhang, Bing Li and Ulf Schlichtmann
Institute for Electronic Design Automation, Technische Universität München, Munich, Germany
{grace-li.zhang, b.li, ulf.schlichtmann}@tum.de

## ABSTRACT

At nanometer manufacturing technology nodes, process variations significantly affect circuit performance. To combat them, post-silicon clock tuning buffers can be deployed to balance timing budgets of critical paths for each individual chip after manufacturing. The challenge of this method is that path delays should be measured for each chip to configure the tuning buffers properly. Current methods for this delay measurement rely on path-wise frequency stepping. This strategy, however, requires too much time from expensive testers. In this paper, we propose an efficient delay test framework (EffiTest) to solve the post-silicon testing problem by aligning path delays using the already-existing tuning buffers in the circuit. In addition, we only test representative paths and the delays of other paths are estimated by statistical delay prediction. Experimental results demonstrate that the proposed method can reduce the number of frequency stepping iterations by more than 94% with only a slight yield loss.

## 1 Introduction

As technology nodes advance, increasing process variations together with aging effects require a nearly unaffordably large timing margin, thus causing expensive overdesign. To combat such challenges post-silicon tuning components and mechanisms have been considered to alleviate the effect of process variations.

A widely used post-silicon tuning technique is clock tuning using delay buffers. For example, the structure of the delay buffer (clock vernier device) in [1] is illustrated in Figure 1. The delay of such a buffer can be adjusted by setting the configuration bits in the three registers. In high-performance designs, these tuning buffers are inserted during the design phase. After manufacturing, the delay values of these buffers are tuned to allot critical paths more timing budget by shifting clock edges toward stages with smaller combinational delays.

In recent years, several methods have already been proposed for statistical timing analysis and optimization of circuits with clock tuning buffers. In [2] a clock scheduling method is developed and clock tuning buffers are selectively inserted to balance the skews due to process variations. In [3] algorithms are proposed to insert buffers into the clock tree to guarantee a given yield, while either the number of buffers or the total area of buffers is minimized. In [4] the yield loss due to process variations and the total cost of clock tuning buffers are formulated together for gate sizing. In [5], the placement of clock tuning buffers is investigated and a considerable benefit is observed when the clock tree is designed using the proposed tuning system. In addition, the work in [6] proposes an efficient post-silicon tuning method by searching a configuration tree combined with graph pruning, and an insertion algorithm to group buffers into clusters. The yield of such a circuit with clock tuning buffers

This work was partly supported by the German Research Foundation (DFG) as part of the Transregional Collaborative Research Centre "Invasive Computing" (SFB/TR 89).

DAC 2016, DOI: 10.1145/2897937.2898017
http://ieeexplore.ieee.org/document/7544303/

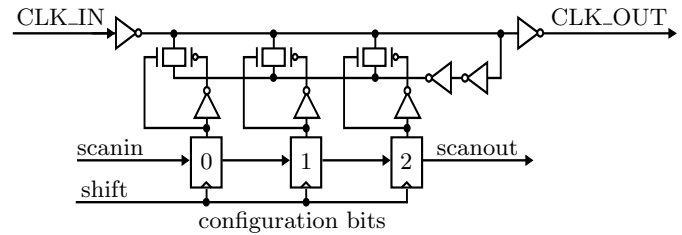

**Figure 1: Post-silicon delay tuning buffer in [1].**

can be evaluated efficiently using the method in [7], and post-silicon testing methods for such circuits have been discussed in [8, 9].

In applying post-silicon tuning buffers, a major challenge is that delays of critical paths need to be measured specifically for each chip after manufacturing. Only with the knowledge of these delays can the tuning buffers be configured properly. However, so far these path delays are still measured using frequency stepping individually [2, 6, 8, 9], which requires much time from an expensive tester.

In this paper, we propose a novel framework (EffiTest) to solve this delay measurement problem. Our contributions are as follows.

- Multiple paths are tested in parallel in our framework. Since we can adjust the existing clock tuning buffers during test, we can align the delays of combinational paths so that a frequency step can capture delay information of multiple paths.
- Instead of exhaustive frequency stepping, we apply statistical delay prediction. With this technique, we need to test only about 10% of the paths whose delays are required for buffer configuration using testers, and the delays of other paths are estimated from the tested delays.
- Experimental results confirm that the number of frequency stepping iterations can be reduced by more than 94%, with only about 2% yield loss.

The rest of this paper is organized as follows. In Section 2 we give an overview of timing constraints for circuits with post-silicon clock tuning buffers and the post-silicon testing problem. We explain the proposed method in detail in Section 3. Experimental results are shown in Section 4. The conclusion is given in Section 5.

## 2 Background of Post-silicon Clock Tuning

In a circuit with post-silicon tuning buffers, the propagation delays of clock paths to flip-flops can be adjusted after manufacturing for each chip individually. The concept of this technique can be explained using the example in Figure 2, where four flip-flops are connected into a loop by combinational paths. The numbers next to inverters represent delays of combinational paths. Although gate delays in advanced technology nodes are statistical [10], they become fixed values after manufacturing.

Without tuning buffers, the minimum clock period of this circuit is 8. If clock edges can be moved by adjusting the delays of the tuning buffers, the minimum clock period can be reduced to 5.5. For example, the buffer value $x_2$ shifts the launching clock edge at F2 2.5 units earlier. Therefore, with a clock period of 5.5, the combinational path between F2 and F3 now has 5.5+2.5=8 time units to finish signal propagation. This shifting of the clock edge reduces the timing budget of the path between F1 and F2 to 5.5-2.5=3 units after post-silicon tuning, which is still sufficient for this path. Note that the buffer delays are defined with respect to a reference clock signal, so that they can have negative values.

The timing imbalance between combinational paths as in Figure 2 potentially appears when process variations become large in advanced technology nodes. For an individual chip, this post-silicon clock tuning is similar to the concept of useful clock skews [11]. The difference is that the tuning values are specific to each individual chip after manufacturing, so that the effect of process variations can be dealt with specifically for each chip.

Timing constraints with clock tuning buffers can be explained using Figure 3, where two flip-flops with such buffers are connected by a combinational circuit. Assume that the clock signal switches at reference time 0. The clock events at flip-flops $i$ and $j$ happen at time $x_i$ and $x_j$, respectively. To meet the setup time and hold time constraints, the following constraints must be satisfied

$$x_i + \overline{d}_{ij} \leq x_j + T - s_j \Longleftrightarrow T \geq D_{ij} + x_i - x_j \quad (1)$$

$$x_i + \underline{d}_{ij} \geq x_j + h_j \Longleftrightarrow x_i - x_j \geq d_{ij} \quad (2)$$

where $x_i$ and $x_j$ are delay values of tuning buffers, $\overline{d}_{ij}$ ($\underline{d}_{ij}$) is the maximum (minimum) delay of the combinational circuit between flip-flops $i$ and $j$, $s_j$ ($h_j$) is the setup (hold) time of flip-flop $j$, $T$ is the clock period, $D_{ij} = \overline{d}_{ij} + s_j$, and $d_{ij} = h_j - \underline{d}_{ij}$. For simplicity, we will still refer to $D_{ij}$ and $d_{ij}$ as path delays in the following discussion.

Owing to area cost, the configurable delay of a clock buffer usually has a limited range. For buffer $i$, this range is specified as

$$r_i \leq x_i \leq r_i + \tau_i \quad (3)$$

where $r_i$ and $\tau_i$ are constants determined by methods such as [3]. In the range (3), $x_i$ may only take discrete values according to the implementation of buffers.

After manufacturing, path delays in chips, e.g. the delays stated next to the inverters in Figure 2, are measured. Thereafter, the configuration values of buffers are determined by finding a feasible solution meeting the constraints (1)–(3).

The most challenging task of using this post-silicon tuning technique, however, is delay measurement of combinational paths in chips after manufacturing. The measured delays should be relatively accurate to configure buffers properly. But the cost due to this delay test must remain low; otherwise, the benefit of using tuning buffers to improve yield may be offset by the ensuing test cost.

In previous methods such as [2, 6, 8, 9], path delays are measured straightforwardly using frequency stepping. In this technique, a path is tested with a given clock period. If the sink flip-flop of this path can latch data correctly, the setup time constraint at the sink flip-flop is met, so that an upper bound of the path delay is found. Thereafter, a smaller clock period is applied until data cannot be latched correctly to find a lower bound of the path delay. With enough frequency steps, the path delay can be approximated by narrowing lower and upper bounds.

Frequency stepping is very easy to use to test path delays, but the number of iterations (frequency steps) might be large if many paths are tested. Though there are some techniques that can be used to combine tests of several paths to reduce the iteration number, no method has considered the fact that the tuning buffers in the circuit can be used to align path delays, so that a clock period can sweep the delay ranges of several paths at the same time. For example, if the buffers in Figure 2 could be preset as shown, one clock period can be used to test these four paths together, because all these paths pass or fail the test at the same time.

To reduce test cost, the correlation information between path delays provided by statistical timing analysis techniques [10] can also be used. Consequently, only a set of representative paths need to be tested while the other path delays are estimated from the test results. In the example in Figure 2, if the correlations between the path delays are high, it is possible that only one or two out of the four paths need frequency stepping.

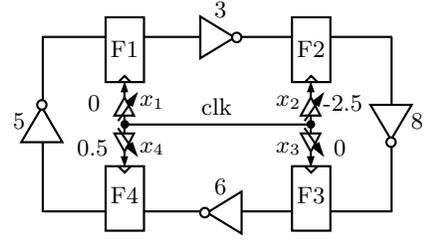

**Figure 2: Post-silicon clock tuning reduces the minimum clock period from 8 to 5.5. Setup time and hold time are assumed as 0.**

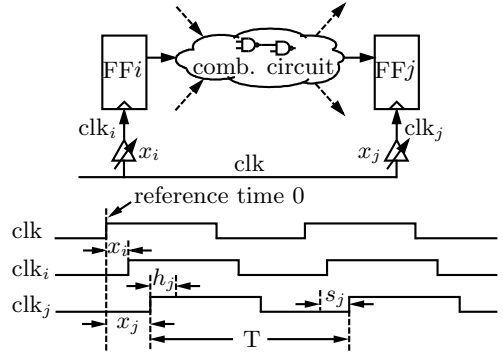

**Figure 3: Timing with tuning buffers.**

## 3 Statistical Prediction and Aligned Delay Test

In this section, we explain our method to reduce the total number of frequency stepping iterations in testing path delays using statistical prediction and delay alignment by tuning buffers. In the test scenario, we assume that the locations of buffers have been determined, using a method such as [3, 12]. We also assume that there is a separate pass/fail test after the buffers are configured, similar to [8]. The flow of the proposed method is summarized in Figure 4.

### 3.1 Statistical Delay Prediction

Consider the test scenario shown in Figure 5, where nodes represent flip-flops and edges represent maximum delays between flip-flops. These maximum delays are needed to configure tuning buffers after manufacturing. Although the number of tuning buffers in the circuit is small, the number of paths that need to be tested may still be large. Consequently, it is impractical to test all these paths with frequency stepping directly, as assumed in [2, 6, 8, 9].

In high-performance designs, the logic gates on a critical path usually are not spread out all over the chip. Therefore, critical paths converging at or leaving from flip-flops with buffers tend to form physical clusters on the chip, as shown in Figure 5. This physical proximity results in high correlation of the path delays [10]. Since a high correlation means that two delays resemble each other in a manufactured chip, actually only a few paths in a highly correlated path set need to be measured in silicon. Thereafter, the delays of other paths can be estimated from these measured delays, using a conditional statistical prediction technique [13], which has been used in [14] to predict the timing performance of a circuit from the measurements of on-chip test structures. This delay prediction technique can be applied to path clusters individually as in Figure 5. In such a cluster, path delays are highly correlated, so that the accuracy of delay prediction can be well maintained.

Assume there are $N$ statistical path delays $\mathbf{D_t}$ which are selected to be measured by frequency stepping, and the delay $d_k$ of another path should be estimated from these $N$ test results $\mathbf{d}_t$. Assume that these delays follow Gaussian distributions. These variables can be written together as $\mathbf{D} = \begin{bmatrix} d_k \\ \mathbf{D}_t \end{bmatrix} \sim \boldsymbol{N}(\boldsymbol{\mu}, \boldsymbol{\Sigma})$ where $\boldsymbol{\mu}$ is the mean value vector of $\mathbf{D}$, $\boldsymbol{\Sigma}$ is the covariance matrix of $\mathbf{D}$, $d_k \sim N(\mu_k, \sigma_k)$ and $\mathbf{D}_t \sim \boldsymbol{N}(\boldsymbol{\mu}_t, \boldsymbol{\Sigma}_t)$. Therefore, $\boldsymbol{\mu}$ and $\boldsymbol{\Sigma}$ can be written as $\boldsymbol{\mu} = \begin{bmatrix} \mu_k \\ \boldsymbol{\mu}_t \end{bmatrix}$, and $\boldsymbol{\Sigma} = \begin{bmatrix} \sigma_k & \boldsymbol{\Sigma}_{k,t} \\ \boldsymbol{\Sigma}_{t,k} & \boldsymbol{\Sigma}_t \end{bmatrix}$, where $\boldsymbol{\Sigma}_{k,t} = \boldsymbol{\Sigma}_{t,k}^T$ is the covariance matrix between $d_k$ and $\mathbf{D}_t$.

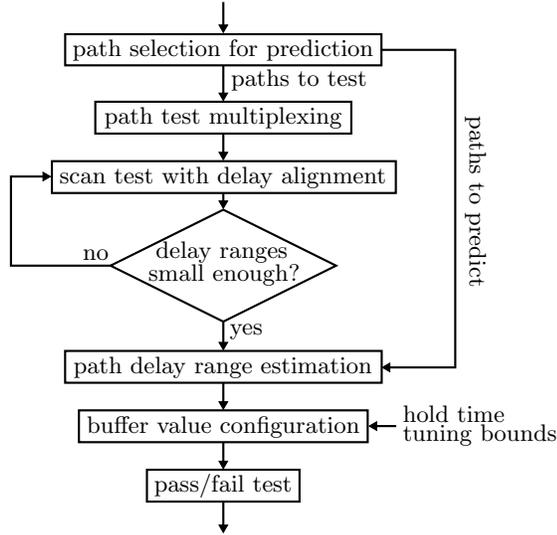

**Figure 4: Test flow for buffer configuration**

By frequency stepping, the delays $\mathbf{D}_t$ can be measured as $\mathbf{d}_t$. According to [13], the mean value $\mu_k$ and the variance $\sigma_k^2$ of the conditional distribution of $d_k$ after $\mathbf{D}_t$ are measured can be expressed as

$$\mu_k' = \mu_k + \boldsymbol{\Sigma}_{k,t}\boldsymbol{\Sigma}_t^{-1}(\boldsymbol{d}_t - \boldsymbol{\mu}_t) \quad (4)$$
$$\sigma_k'^2 = \sigma_k^2 - \boldsymbol{\Sigma}_{k,t}\boldsymbol{\Sigma}_t^{-1}\boldsymbol{\Sigma}_{t,k}. \quad (5)$$

Since the second product term in (5) is positive, the variance of the delay $d_k$ can be reduced. Consequently, the value of $d_k$ is limited into a narrow range and it may become unnecessary to measure the exact delay of $d_k$ for buffer configuration if the correlation between $d_k$ and $\mathbf{D}_t$ is high. On the other hand, a small correlation allows the delays to vary freely, leading to a relatively large variance even after statistical prediction is applied.

In the discussion above, all variables are assumed as Gaussian. This is an assumption widely used in statistical timing analysis [10]. The proposed method, however, only requires an estimation of the upper bounds of delays for buffer configuration (described in later sections), so that the exact distributions do not affect the result much. For a non-Gaussian distribution, independent component analysis (ICA) may also be considered as in [15] with an expansion on conditional distribution.

Since the quality of delay prediction relies on the magnitude of correlation, we partition path delays in different groups. We first extract the paths with high correlations. These paths have a good delay prediction accuracy, and only a small number of paths from this group need to be tested. We then lower the correlation threshold to extract further path groups until all paths are extracted. This grouping technique can handle the case that there are several clusters of critical paths that are far away in the circuit. The correlations between paths from different clusters may be small, but inside each cluster the correlation is still high.

For each path group, we decompose the delays using principal component analysis (PCA) [16, 17] to identify the principal components (PCs) shared by correlated variables. Since only the PCs carry correlation information and only those are useful in predicting path delays, we need only to select those paths that can capture the values of these PCs by frequency stepping. Assuming that the number of PCs in the $i$th path group $P_i$ is $|PC_i|$, we select the same number ($|PC_i|$) of paths from $P_i$ [14]. After decomposition, the delays of paths $P_i$ are represented as linear combinations of PCs. We first select the path with the largest coefficient for the first PC. Thereafter, from the remaining paths, we select the one with the largest coefficient for the next PC. This process is repeated until

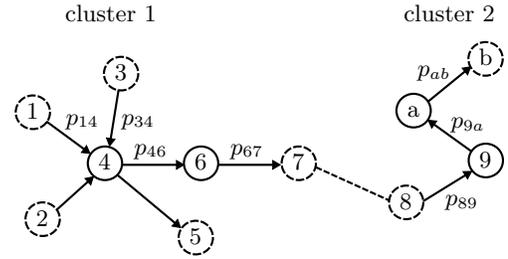

**Figure 5: Test scenario.** The nodes with solid (dashed) lines represent flip-flops with (without) tuning buffers.

$|PC_i|$ paths are selected.

The pseudocode of path grouping and path selection for frequency test is shown in Procedure 1.

| **Procedure 1:** Select Paths |
|---|
| **1**  $P$: paths whose delays are required. |
| **2**  $P_t$: paths to be tested. |
| **3**  $corr\_th$=0.95; $P_t = \varnothing$; $i = 1$; |
| **4**  **while** $P \neq \varnothing$ **do** |
| **5**    $P_i$=extract_paths($P$, $corr\_th$); |
| **6**    $\Sigma_i$=cov($P_i$); |
| **7**    $PC_i$=PCA($\Sigma_i$); |
| **8**    $P_{t,i}$=select_paths($P_i$, $PC_i$);       // $|P_{t,i}| = |PC_i|$ |
| **9**    $P_t = P_t \bigcup P_{t,i}$; |
| **10**   $corr\_th = corr\_th - 0.05$; |
| **11**   $i = i + 1$; |
| **12** **end while** |

### 3.2 Path Test Multiplexing

Since the individual delays of selected paths ($P_t$ in Procedure 1) should be measured, paths converging at or leaving from the same flip-flop cannot be tested in parallel. For example, the paths $p_{14}$ and $p_{34}$ in Figure 5 cannot be tested together, because a data latching failure at node 4 cannot be identified as a failure of either $p_{14}$ or $p_{34}$ definitely. Consequently, paths that are measured together should be arranged in series. For example, paths $p_{14}$, $p_{46}$, $p_{67}$, $p_{89}$, $p_{9a}$, $p_{ab}$ can be tested with the same clock period together. These paths are called a *batch* in the following discussion. In real circuits, there might be cases that some paths in a test batch cannot be activated by ATPG vectors at the same time due to logic masking. These paths can be set as mutually exclusive and arranged into different test batches.

Since the delays of paths in a test batch can be measured in parallel, naturally we should arrange paths to be tested into as few batches as possible to reduce the overall number of frequency stepping iterations. This arrangement can be determined easily using a depth-first search or a simple ILP model so that we skip the detailed discussion here.

After the test batches are formed, there might still be some unoccupied slots in a test batch because paths might not be distributed evenly at flip-flops with buffers. Since the path batches should be tested anyway, we add additional paths to these empty test slots to gather more delay information. According to (5), the variance of a path after estimation does not rely on the results of delay test $d_t$. Since a large variance means that the delay cannot be estimated with enough accuracy, we add such paths with large variances to the empty slots in the identified test batches so that their delays can also be measured to reduce the predicted delay ranges.

### 3.3 Test with Delay Alignment by Tuning Buffers

After path batches are identified, they should be tested using frequency stepping to determine their path delays. In this section, we discuss how the delays of paths in a single batch are measured. Note this is the only step in the proposed framework that is executed by expensive testers that are able to generate various clock signals with a high accuracy.

In frequency stepping, a clock period is applied to the chip under test and the paths in a test batch are sensitized by test vectors.

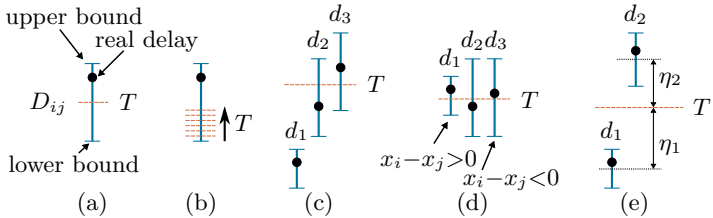

**Figure 6: Frequency stepping and delay range alignment.**

If the setup time constraint (1) at a flip-flop is violated, the data at this flip-flop cannot be latched correctly. This error shows that $D_{ij} + x_i - x_j$ is larger than $T$ so that $T$ is its lower bound. On the other hand, if the clock period is large enough so that there is no timing violation, the constraint (1) is met and $T$ is an upper bound of $D_{ij} + x_i - x_j$. By applying different clock periods in a binary search style, the value of $D_{ij}$ can be approximated with a given accuracy.

Consider the case shown in Figure 6a, where a delay has given upper and lower bounds. These bounds are initialized with $\mu \pm 3\sigma$, where $\mu$ and $\sigma$ are the mean value and the standard deviation of the delay calculated by statistical timing analysis. When the delay is tested with a given clock period $T$ in an iteration, either a new upper bound or a new lower bound of it is generated. Consequently, the corresponding delay range is partitioned into two parts by $T$ and the real delay value falls into one of them. To partition the delay range efficiently, it is preferable that $T$ is aligned to the center (middle point) of the range. Otherwise, $T$ might not partition the delay range evenly, but instead slices it in small steps, leading to many test iterations to estimate the delay, as illustrated in Figure 6b.

When several path delays in one test batch are considered as in Figure 6c, it is not always possible to partition all the delay ranges evenly with one clock period. However, we can still find a clock period $T$ that partitions several delay ranges at the same time, so that the ranges of these delays can be reduced in one test iteration.

To use a clock period $T$ to partition multiple delay ranges, there must be some overlap between the delay ranges, such as $d_2$ and $d_3$ in Figure 6c. According to (1), the actual constraint that is tested using $T$ is $D_{ij}+x_i-x_j$. Since the tuning buffers are already deployed in the circuit and their values $x_i$ and $x_j$ can be adjusted through the scan chain, we change the value of $x_i - x_j$ to align the delay ranges, as illustrated in Figure 6d. Consequently, a clock period can partition more delay ranges so that the delays can be measured more efficiently compared with the case in Figure 6c. Because the configuration bits of buffers can be scanned into the chip under test together with the test vectors, this technique requires no change to the existing test platform.

In real circuits, the buffer values $x_i$ and $x_j$ can only be adjusted in a limited range as specified by (3). In addition, these buffer values may affect more than one path delay. For example, in Figure 5 the buffer value of node 4 affects all the paths converging at or leaving from it. To test the path delays efficiently, we need to find a proper set of buffer values to align the ranges of path delays as much as possible.

Assume that the upper and lower bounds of $D_{ij}$ between nodes $i$ and $j$ are $u_{ij}$ and $l_{ij}$, respectively. When the buffers at the source and sink nodes of the path are considered, the lower bounds and the upper bounds are shifted by $x_i - x_j$ as defined in (1). Therefore, the distance $\eta_{ij}$ between a given $T$ and the center of the shifted range of the path delay $D_{ij}$ can be expressed as

$$\eta_{ij} = |T - ((u_{ij} + l_{ij})/2 + x_i - x_j)|. \tag{6}$$

If we minimize the sum of $\eta_{ij}$ from all delay ranges, the resulting $T$ will approximate the centers of delay ranges as much as possible, while the buffer values $x_i$ and $x_j$ are also determined.

Minimizing the sum of $\eta_{ij}$ directly, however, cannot handle the special case in Figure 6e where the two delay ranges still do not overlap even after the buffer values have been adjusted to the limit.

In this case, the sum of distances $\eta_1 + \eta_2$ is independent of where $T$ is placed between the centers of the two ranges. To solve this problem, we sort the centers of delay ranges determined in the previous test iteration. Thereafter, we assign the weight $k_0$ to the range whose center is in the middle of the sorted list, and reduce the weights of other ranges by $k_d$ successively. In the proposed method, we set $k_0 \gg k_d$, so that the ranges at the middle of the sorted list have slightly higher priorities. With this weight assignment, the weights of the two ranges in Figure 6e are different so that the next test clock period $T$ should align at the center of the range with the larger weight.

The optimization problem to determine the clock period $T$ and the corresponding set of buffer values $x_i$ and $x_j$ to align delay ranges can thus be expressed as

$$\text{minimize} \quad \sum_{i,j} k_{ij} \eta_{ij} \tag{7}$$

subject to $\forall$ path $p_{ij}$ in the test batch
$$T - ((u_{ij} + l_{ij})/2 + x_i - x_j) \leq \mathcal{M} z_{ij}^p \tag{8}$$
$$(T - ((u_{ij} + l_{ij})/2 + x_i - x_j)) - \eta_{ij} \leq \mathcal{M}(1 - z_{ij}^p) \tag{9}$$
$$-(T-((u_{ij} + l_{ij})/2 + x_i - x_j)) + \eta_{ij} \leq \mathcal{M}(1 - z_{ij}^p) \tag{10}$$
$$-(T-((u_{ij} + l_{ij})/2 + x_i - x_j)) \leq \mathcal{M} z_{ij}^n \tag{11}$$
$$-(T-((u_{ij} + l_{ij})/2 + x_i - x_j)) - \eta_{ij} \leq \mathcal{M}(1 - z_{ij}^n) \tag{12}$$
$$(T - ((u_{ij} + l_{ij})/2 + x_i - x_j)) + \eta_{ij} \leq \mathcal{M}(1 - z_{ij}^n) \tag{13}$$
$$r_i \leq x_i \leq r_i + \tau_i, \; r_j \leq x_j \leq r_j + \tau_j \tag{14}$$

where (8)–(13) are linear constraints transformed from (6) and $\mathcal{M}$ is a very large positive constant [18]; $z_{ij}^p$ and $z_{ij}^n$ are two 0-1 variables corresponding to the two cases that $T - ((u_{ij} + l_{ij})/2 + x_i - x_j)$ are no less than zero and no greater than zero, respectively. (14) defines the ranges of buffer values as in (3).

After the clock frequency and the corresponding buffer values are determined by solving the ILP problem (7)–(14), the paths in the current batch are tested. According to the test result, either the upper bounds or the lower bounds of their delays are updated. If the distance between the range bounds $u_{ij}$ and $l_{ij}$ of a path is smaller than a threshold $\epsilon$, the path is removed from the current batch. The test iterations finish when all paths in the batch have been removed. The pseudocode of the test process is shown in Procedure 2.

**Procedure 2:** Delay Test

```
1   B: the queue of test batches.
2   while B ≠ ∅ do
3       B_k=pop_top(B);
4       while B_k contains an edge do
5           T=cmp_freq(B_k);                    // solve (7)-(14)
6           test_with_frequency_stepping(B_k, T);
7           foreach p_ij in B_k do
8               if passed(p_ij) then
9                   u_ij=T - x_i + x_j;
10              else
11                  l_ij=T - x_i + x_j;
12              end if
13              if u_ij - l_ij < ε then
14                  remove_edge(p_ij, B_k);
15              end if
16          end foreach
17      end while
18  end while
```

### 3.4 Buffer Configuration with Delay Estimation

After a path in $P_t$ is tested by frequency stepping, its delay has been in a range with a lower bound and an upper bound. For another delay $d_k$ that is not measured directly but to be estimated, (4) and (5) are used to calculate the mean value $\mu'_k$ and the standard deviation $\sigma'_k$. According to (4) and (5), $\sigma'_k$ is determined exclusively by the covariance matrix, but $\mu'_k$ is affected by $\boldsymbol{d}_t$, which are the delays measured by frequency stepping. When calculating $\mu'_k$, we use the upper bounds of $\boldsymbol{d}_t$ so that the estimated delays are conservative.

Table 1: Test Results With Delay Alignment and Statistical Prediction

| Circuit | | | Our Method | | | | Comparison | | | | Runtime | | |
|---|---|---|---|---|---|---|---|---|---|---|---|---|---|
| | $n_s$ | $n_g$ | $n_b$ | $n_p$ | $n_{p_t}$ | $t_a$ | $t_v$ | $t'_a$ | $t'_v$ | $r_a(\%)$ | $r_v(\%)$ | $T_p(s)$ | $T_t(s)$ | $T_s(s)$ |
| s9234 | 211 | 5597 | 2 | 80 | 15 | 37 | 2.47 | 700 | 8.75 | 94.71 | 71.77 | 6.58 | 0.09 | 0.00 |
| s13207 | 638 | 7951 | 5 | 485 | 19 | 39 | 2.05 | 4001 | 8.25 | 99.03 | 75.15 | 16.75 | 0.06 | 0.00 |
| s15850 | 534 | 9772 | 5 | 397 | 22 | 76 | 3.45 | 3684 | 9.28 | 97.94 | 62.82 | 50.51 | 0.17 | 0.01 |
| s38584 | 1426 | 19253 | 7 | 370 | 21 | 62 | 2.95 | 3093 | 8.36 | 98.00 | 64.71 | 90.45 | 0.15 | 0.01 |
| mem_ctrl | 1065 | 10327 | 10 | 3016 | 62 | 195 | 3.15 | 27415 | 9.09 | 99.29 | 65.35 | 622.63 | 0.36 | 0.02 |
| usb_funct | 1746 | 14381 | 17 | 482 | 32 | 114 | 3.56 | 4569 | 9.48 | 97.51 | 62.45 | 118.48 | 0.17 | 0.02 |
| ac97_ctrl | 2199 | 9208 | 21 | 780 | 78 | 288 | 3.69 | 7340 | 9.41 | 96.08 | 60.79 | 81.63 | 0.30 | 0.01 |
| pci_bridge32 | 3321 | 12494 | 32 | 3472 | 84 | 298 | 3.55 | 29061 | 8.37 | 98.97 | 57.59 | 749.31 | 1.19 | 1.59 |

Since the variances of estimated delays are often not zero, indicating that purely random variations still affect path delays, we assign a lower bound and an upper bound $\mu'_k - 3\sigma'_k$ and $\mu'_k + 3\sigma'_k$ for an estimated delay, so that all path delays are constrained similarly for the following buffer configuration.

A real delay may take any value in the range defined by the lower and upper bounds, but the exact location of this delay in the range is unknown due to test resolution and delay estimation. In this situation, a conservative method to configure the buffers is to assume the upper bounds of the ranges as path delays, so that the chip always works with the resulting buffer configuration. This method, however, may incorrectly report some chips as nonfunctional due to this delay overestimation. To solve this problem, we try to find a buffer configuration for a chip while assuming the delays are close to their corresponding upper bounds as much as possible. By minimizing the distance of the assumed delays from their corresponding upper bounds when determining the buffer configuration, the chance that the chip works after configuration becomes large, so that the final pass/fail test will accept most post-silicon configured chips as functional.

The optimization problem to find a buffer configuration while minimizing the distance $\xi$ of the assumed delays from the corresponding upper bounds is described as follows.

$$\text{minimize} \quad \xi \qquad (15)$$
$$\text{subject to} \quad \forall \text{ path } p_{ij}$$
$$T_d \geq D'_{ij} + x_i - x_j \qquad (16)$$
$$l_{ij} \leq D'_{ij} \leq u_{ij}, \xi \geq u_{ij} - D'_{ij} \qquad (17)$$
$$r_i \leq x_i \leq r_i + \tau_i, r_j \leq x_j \leq r_j + \tau_j \qquad (18)$$

where $D'_{ij}$ is the assumed delay value of a path during buffer configuration; $T_d$ is the designated clock period for the design; (16) and (18) are derived from (1) and (3), respectively. By solving the optimization problem (15)–(18), a set of buffer configuration values $x_i$ and $x_j$ can be found.

### 3.5 Tuning Bounds due to Hold Time Constraints

In the discussion above, we do not consider hold time constraints. However, tuning buffers may affect hold time constraints significantly if they are configured improperly. For example, in Figure 3, if $x_j$ is much larger than $x_i$, the constraint (2) may be violated.

As shown in (2), hold time constraints are affected by $x_i - x_j$ instead of individual values of $x_i$ and $x_j$. In our method, we do not test against hold time violations after configuring buffers. Instead, we set a lower bound $\lambda_{ij}$ for $x_i - x_j$ by sampling the statistical distribution of $d_{ij}$ in (2) so that a given yield can be maintained.

Consider the case that $d_{ij}$ in (2) is sampled $M$ times for all short paths and its value in the $k$th sample is $d_{ij,k}$. For the $k$th sample, we use a 0-1 variable $y_k$ to represent that the lower bound $\lambda_{ij}$ meet

$$\lambda_{ij} - d_{ij,k} \geq \mathcal{M}(y_k - 1), \quad \text{for all short paths } p_{ij} \qquad (19)$$

where $\mathcal{M}$ is a very large constant. The yield of the circuit with respect to hold time can thus be constrained as

$$\sum y_i / M \geq Y, \quad i = 1, 2, \ldots M \qquad (20)$$

where $Y$ is a given yield for hold time constraints, set to 0.99 in our method. To allow buffers to have the largest freedom in value configuration, we minimize the sum of all the lower bounds $\sum_{i,j} \lambda_{ij}$.

After $\lambda_{ij}$ are determined, the buffer configuration values can be constrained to avoid hold time violation, as

$$x_i - x_j \geq \lambda_{ij}. \qquad (21)$$

This constraint is added into the optimization problems in Section 3.3 and Section 3.4 to incorporate hold time constraints to determine buffer values $x_i$ and $x_j$.

## 4 Experimental Results

The proposed framework was implemented in C++ and tested using a 3.20 GHz CPU. We demonstrate the results with circuits from ISCAS89 and TAU13 benchmark sets. Information about these circuits is shown in Table 1, where $n_s$ is the number of flip-flops and $n_g$ the number of logic gates. The number of inserted tuning buffers was less than 1% of the number of flip-flops. The numbers of buffers are shown in the column $n_b$. As in [19], we assumed that the maximum allowed buffer ranges were 1/8 of the original clock period and all tuning delays were set to be discrete with 20 steps. The logic gates in the circuits were mapped to a library from an industry partner. The standard deviations of transistor length, oxide thickness and threshold voltage were set to 15.7%, 5.3% and 4.4% of the nominal values. The correlation of variations in two side-by-side gates was set to 1 and the correlation due to global variations was set to 0.25. The ILP solver for the optimization problems was Gurobi [20].

In Table 1 the column $n_p$ shows the numbers of paths whose delays are required for buffer configuration. Although there are only a small number ($n_b$) of buffers in the circuits, the numbers of paths to be tested ($n_p$) are still large, specially for the circuits mem_ctrl and pci_bridge32. The column $n_{p_t}$ shows the numbers of paths that are actually tested by the proposed method. Due to statistical prediction, only a small number of paths were selected so that the number of test iterations can be reduced directly. In our experiments, we tested 10 000 simulated chips. The column $t_a$ shows the average number of frequency stepping iterations for each chip using the proposed method, and the column $t_v$ shows the average number of iterations per path, where $t_v = t_a/n_{p_t}$.

For comparison, we implemented the method applying frequency stepping to each path individually, as assumed in [2, 6, 8, 9]. The column $t'_a$ in Table 1 shows the total numbers of test iterations. Since there are a lot of paths that should be tested ($n_p$), $t'_a$ are extraordinarily large. These numbers confirm that the straightforward frequency stepping method is impractical for large circuits. Furthermore, the column $t'_v$ shows the average numbers of frequency stepping iterations per path, where $t'_v = t'_a/n_p$. Comparing the columns $t_v$ and $t'_v$, we can find that the proposed method is much more efficient, due to the test multiplexing technique described in Section 3.2 and the aligned test technique described in Section 3.3. The columns $r_a(\%)$ and $r_v(\%)$ show the reduction ratios of the test iterations per chip and the test iterations per path, where $r_a = (t'_a - t_a)/t'_a * 100$ and $r_v = (t'_v - t_v)/t'_v * 100$. Combining statistical prediction and aligned delay test, the overall test effort can be reduced by more than 94% (94.71%~99.29%). If we look at the ratios of test iterations per path ($r_v(\%)$), we can find that the test reductions are between 57.59% and 75.15%. This reduction comes only from test multiplexing and aligned delay test, while the statistical prediction technique does not affect this ratio much. Both comparisons, however, confirm that the proposed test framework reduces test cost

Table 2: Yield Comparison

| Circuit | $T_1$ | | | $T_2$ | | |
|---|---|---|---|---|---|---|
| | $y_i(\%)$ | $y_t(\%)$ | $y_r(\%)$ | $y_i(\%)$ | $y_t(\%)$ | $y_r(\%)$ |
| s9234 | 77.11 | 75.80 | 1.31 | 95.94 | 95.61 | 0.33 |
| s13207 | 72.37 | 72.09 | 0.28 | 96.42 | 96.03 | 0.39 |
| s15850 | 69.34 | 69.09 | 0.25 | 94.33 | 94.10 | 0.23 |
| s38584 | 85.97 | 85.01 | 0.96 | 98.48 | 97.10 | 1.38 |
| mem_ctrl | 67.11 | 64.98 | 2.13 | 94.58 | 92.40 | 2.18 |
| usb_funct | 71.77 | 69.40 | 2.37 | 96.57 | 94.60 | 1.97 |
| ac97_ctrl | 75.05 | 73.40 | 1.65 | 94.92 | 93.09 | 1.83 |
| pci _bridge32 | 73.66 | 71.50 | 2.16 | 96.76 | 95.71 | 1.05 |

significantly.

The runtimes of the proposed method are shown in the last three columns in Table 1, where $T_p$ is the runtime for path grouping and selection, test multiplexing and hold time bound computation. Because these steps are performed offline, the runtime is already acceptable. The column $T_t(s)$ shows the average runtime when computing the clock period $T$ and the buffer configuration values for all test batches of a chip. Since this computation can be performed in parallel while path batches are tested, the runtime is also acceptable compared with the execution time of scan test. The last column $T_s(s)$ shows the runtime to determine the final buffer values using the method in Section 3.4. This step is not performed on expensive testers so that the efficiency is good enough.

In the proposed framework, the results of aligned delay test produce lower and upper bounds for delays. This inaccuracy cannot be avoided due to the nature of delay test and it affects the yields of the circuits after buffer configuration. In addition, the technique of statistical prediction also introduces configuration inaccuracy in the estimated delays. Consequently, it is expected that the yields of the circuits should drop from the ideal yields with delays measured exactly. We tested several cases with two clock periods $T_1$ and $T_2$ and the results are shown in Table 2. For $T_1$ and $T_2$ the original yields without buffers were 50% and 84.13%, respectively. The column $y_i$ shows the yields with a perfect delay measurement; the column $y_t$ shows the yields with delays measured by the proposed method; and the column $y_r$ shows the yield drops due to the inaccuracy in the tested delays, where $y_r = y_i - y_t$. In these results, we can see that the yield drops are around 1-2%, where the improved yields are still far better than the yields without buffers, 50% and 84.13%, respectively.

Since the results of the statistical prediction technique in Section 3.1 depend on the correlations between path delays, we manually increased the standard deviations of all delays by 10%. Since we did not change the covariance matrix between variables, this change led to a large increase in the purely random parts of the delays. Figure 7 shows the yield results of three cases: 1) no buffers in the circuits; 2) with buffers and the buffer configurations generated by the proposed method; 3) with buffers and perfect buffer configurations. The latter two cases clearly demonstrate that the yields were still improved impressively due to tuning buffers. When testing and configuring the buffer values with the proposed method, the yields dropped more from the ideal case than the cases in Table 1 due to the increased random variation. But the final results are still good considering the significant reduction in test cost.

To verify the effectiveness of test multiplexing and aligned delay ranges described in Section 3.2 and Section 3.3, we applied them directly to reduce test iterations without statistical prediction. Figure 8 shows the comparison of the numbers of test iterations per path in three cases: 1) path-wise frequency stepping; 2) test multiplexing without delay alignment using buffers; 3) multiplexing with delay alignment using buffers (the proposed method). The second case uses the method in Section 3.2 and Section 3.3, but all the buffers values were set to zero. Comparing the results of the first case and the second case, we can see that test multiplexing is a powerful technique to reduce test iterations. When the technique of delay alignment is applied, test iterations can be reduced further, as demonstrated by the third case. These results confirm that even without taking advantage of the correlations between path delays, the proposed method can still reduce test cost significantly.

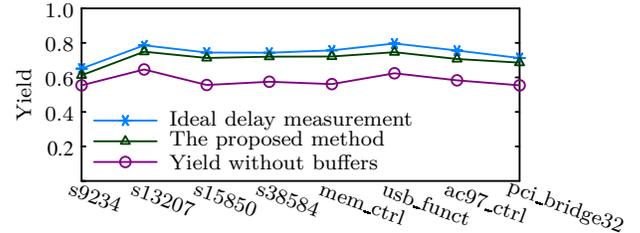

Figure 7: Yield with enlarged random variation

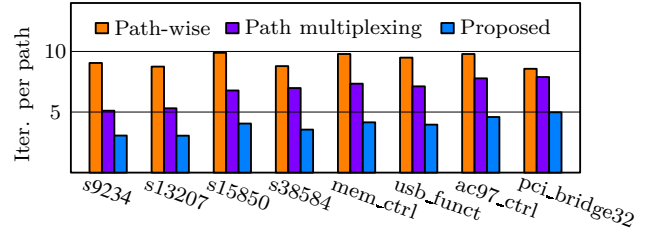

Figure 8: Test comparison without statistical prediction

## 5 Conclusion

In this paper we propose an efficient framework to reduce test cost in configuring tuning buffers in high-performance designs. This framework combines the techniques statistical prediction and aligned delay test with path multiplexing, with which the number of test iterations can be reduced by more than 94%. The effectiveness of these techniques has been confirmed by experimental results using ISCAS89 and TAU13 benchmark circuits.

## 6 References

[1] S. Naffziger, B. Stackhouse, T. Grutkowski, D. Josephson, J. Desai, E. Alon, and M. Horowitz. The implementation of a 2-core, multi-threaded Itanium family processor. *IEEE J. Solid-State Circuits*, 41(1):197–209, January 2006.
[2] J. Tsai, D. Baik, C. C.-P. Chen, and K. K. Saluja. A yield improvement methodology using pre- and post-silicon statistical clock scheduling. In *Proc. Int. Conf. Comput.-Aided Des.*, pages 611–618, 2004.
[3] J. Tsai, L. Zhang, and C. C.-P. Chen. Statistical timing analysis driven post-silicon-tunable clock-tree synthesis. In *Proc. Int. Conf. Comput.-Aided Des.*, pages 575–581, 2005.
[4] V. Khandelwal and A. Srivastava. Variability-driven formulation for simultaneous gate sizing and post-silicon tunability allocation. In *Proc. Int. Symp. Phys. Des.*, pages 11–18, 2007.
[5] K. Nagaraj and S. Kundu. A study on placement of post silicon clock tuning buffers for mitigating impact of process variation. In *Proc. Design, Autom., and Test Europe Conf.*, pages 292–295, 2009.
[6] Z. Lak and N. Nicolici. A novel algorithmic approach to aid post-silicon delay measurement and clock tuning. *IEEE Trans. Comput.*, 63(5):1074–1084, May 2014.
[7] B. Li, N. Chen, and U. Schlichtmann. Fast statistical timing analysis for circuits with post-silicon tunable clock buffers. In *Proc. Int. Conf. Comput.-Aided Des.*, pages 111–117, 2011.
[8] K. Nagaraj and S. Kundu. An automatic post silicon clock tuning system for improving system performance based on tester measurements. In *Proc. Int. Test Conf.*, pages 1–8, 2008.
[9] D. Tadesse, J. Grodstein, and R. I. Bahar. AutoRex: An automated post-silicon clock tuning tool. In *Proc. Int. Test Conf.*, pages 1–10, 2009.
[10] D. Blaauw, K. Chopra, A. Srivastava, and L. Scheffer. Statistical timing analysis: from basic principles to state of the art. *IEEE Trans. Comput.-Aided Design Integr. Circuits Syst.*, 27(4):589–607, April 2008.
[11] J.P. Fishburn. Clock skew optimization. *IEEE Trans. Comput.*, 39(7):945–951, July 1990.
[12] Grace Li Zhang, Bing Li, and Ulf Schlichtmann. Sampling-based buffer insertion for post-silicon yield improvement under process variability. In *Proc. Design, Autom., and Test Europe Conf.*, 2016.
[13] R. A. Johnson and D. W. Wichern. *Applied multivariate statistical analysis*. Pearson Prentice Hall, Upper Saddle River, 2007.
[14] Q. Liu and S. S. Sapatnekar. A framework for scalable postsilicon statistical delay prediction under process variations. *IEEE Trans. Comput.-Aided Design Integr. Circuits Syst.*, 28(8):1201–1212, August 2009.
[15] J. Singh and S. S. Sapatnekar. A scalable statistical static timing analyzer incorporating correlated non-gaussian and gaussian parameter variations. *IEEE Trans. Comput.-Aided Design Integr. Circuits Syst.*, 27(1):160–173, January 2008.
[16] I.T. Jolliffe. *Principal Component Analysis*. Springer, 2002.
[17] H. Chang and S. S. Sapatnekar. Statistical timing analysis under spatial correlations. *IEEE Trans. Comput.-Aided Design Integr. Circuits Syst.*, 24(9):1467–1482, September 2005.
[18] D.S. Chen, R.G. Batson, and Y. Dang. *Applied Integer Programming: Modeling and Solution*. Wiley, 2011.
[19] S. Tam, S. Rusu, U. Nagarji Desai, R. Kim, Ji Zhang, and I. Young. Clock generation and distribution for the first IA-64 microprocessor. *IEEE J. Solid-State Circuits*, 35(11):1545–1552, November 2000.
[20] Gurobi Optimization, Inc. Gurobi optimizer reference manual, 2013.